\documentclass[%
prl,
twocolumn,
superscriptaddress,
 amsmath,amssymb,
]{revtex4-1}

\usepackage{graphicx}% Include figure files
\usepackage{dcolumn}% Align table columns on decimal point
\usepackage{bm}% bold math
\usepackage{siunitx}
\usepackage{xcolor}
\begin{document}

%\preprint{APS/123-QED}
\title{Time-resolved diffraction and photoelectron spectroscopy investigation of the reactive molecular beam epitaxy of Fe\textsubscript{3}O\textsubscript{4} ultrathin films}% Force line breaks with \\
\author{\underline{Tobias Pohlmann}}
\email{tobias.pohlmann@desy.de}
\affiliation{Deutsches Elektronen-Synchrotron DESY, Notkestr. 85, 22607 Hamburg, Germany}
\affiliation{Department of Physics, Osnabr\"uck University, 49076 Osnabr\"uck, Germany}
\author{Martin Hoppe}
\affiliation{Deutsches Elektronen-Synchrotron DESY, Notkestr. 85, 22607 Hamburg, Germany}
\affiliation{Department of Physics, Osnabr\"uck University, 49076 Osnabr\"uck, Germany}
\author{Jannis Thien}
\affiliation{Department of Physics, Osnabr\"uck University, 49076 Osnabr\"uck, Germany}
\author{Arka Bikash Dey}
\affiliation{Deutsches Elektronen-Synchrotron DESY, Notkestr. 85, 22607 Hamburg, Germany}
\author{Andreas Alexander}
\affiliation{Department of Physics, Osnabr\"uck University, 49076 Osnabr\"uck, Germany}
\author{Kevin Ruwisch}
\affiliation{Department of Physics, Osnabr\"uck University, 49076 Osnabr\"uck, Germany}
\author{Olof Gutowski}
\affiliation{Deutsches Elektronen-Synchrotron DESY, Notkestr. 85, 22607 Hamburg, Germany}
\author{Jan R\"oh}
\affiliation{Deutsches Elektronen-Synchrotron DESY, Notkestr. 85, 22607 Hamburg, Germany}
\author{Andrei Gloskovskii}
\affiliation{Deutsches Elektronen-Synchrotron DESY, Notkestr. 85, 22607 Hamburg, Germany}
\author{Christoph Schlueter}
\affiliation{Deutsches Elektronen-Synchrotron DESY, Notkestr. 85, 22607 Hamburg, Germany}
\author{Karsten K\"upper}
\affiliation{Department of Physics, Osnabr\"uck University, 49076 Osnabr\"uck, Germany}
\author{Joachim Wollschl\"ager}
\affiliation{Department of Physics, Osnabr\"uck University, 49076 Osnabr\"uck, Germany}
\author{Florian Bertram}
\affiliation{Deutsches Elektronen-Synchrotron DESY, Notkestr. 85, 22607 Hamburg, Germany}
\date{\today}% It is always \today, today,
             %  but any date may be explicitly specified             
                      
\begin{abstract}
We present time-resolved high energy x-ray diffraction (tr-HEXRD), time-resolved hard x-ray photoelectron spectroscopy (tr-HAXPES) and time-resolved grazing incidence small angle x-ray scattering (tr-GISAXS) data of the reactive molecular beam epitaxy (RMBE) of $\mathrm{Fe_3O_4}$ ultrathin films on various substrates. Reciprocal space maps are recorded during the deposition of $\mathrm{Fe_3O_4}$ on $\mathrm{SrTiO_3(001)}$, MgO(001) and NiO/MgO(001) in order to observe the temporal evolution of Bragg reflections sensitive to the octahedral and tetrahedral sublattices of the inverse spinel structure of $\mathrm{Fe_3O_4}$. A time delay between the appearance of rock salt and spinel-exclusive reflections reveals that first, the iron oxide film grows with Fe\textsubscript{1-$\delta$}O rock salt structure with exclusive occupation of octahedral lattice sites.  When this film is $1.1\,$nm thick, the further growth of the iron oxide film proceeds in the inverse spinel structure, with both octahedral and tetrahedral lattice sites being occupied. In addition, iron oxide on $\mathrm{SrTiO_3(001)}$ initially grows with none of these structures. Here, the formation of the rock salt structure starts when the film is $1.5\,$nm thick. This is confirmed by tr-HAXPES data obtained during growth of iron oxide on $\mathrm{SrTiO_3(001)}$, which demonstrate an excess of $\mathrm{Fe^{2+}}$ cations in growing films thinner than $3.2\,$nm.
This rock salt phase only appears during growth and vanishes after the supply of the Fe molecular beam is stopped. Thus, it can be concluded the rock salt structure of the interlayer is a property of the dynamic growth process. The tr-GISAXS data link these structural results to an island growth mode of the first $2-3\,$nm on both MgO(001) and $\mathrm{SrTiO_3(001)}$ substrates.
\end{abstract}
\pacs{Valid PACS appear here}% PACS, the Physics and Astronomy
                             % Classification Scheme.
%\keywords{Suggested keywords}%Use showkeys class option if keyword
                              %display desired
\maketitle

%\tableofcontents
\section{\label{sec:intro}Introduction}
Iron oxides grow in a variety of crystal structures and stoichiometries, and their electronical and magnetic properties vary significantly between these phases.
Both hematite ($\alpha$-$\mathrm{Fe_2O_3}$) and maghemite ($\gamma$-$\mathrm{Fe_2O_3}$) are semiconductors and represent the iron oxides with the highest oxidation state, with all iron ions occuring in the $\mathrm{Fe^{3+}}$ charge state \cite{Litter92,Grau-Crespo10}. $\alpha$-$\mathrm{Fe_2O_3}$ crystallizes in a corundum structure \cite{Pauling25} and is a canted antiferromagnet at room temperature with a N\'eel temperature $T_\mathrm{N}=683^\circ$C \cite{Cornell03}, while $\gamma$-$\mathrm{Fe_2O_3}$ is ferrimagnet with a Curie temperature of $T_\mathrm{C}=620^\circ$C \cite{Gehring09} and grows in a defect spinel structure \cite{Cornell03}.
Magnetite ($\mathrm{Fe_3O_4}$) is a half-metallic ferrimagnet with a Curie temperature of $T_\mathrm{C}=580^\circ$C. It crystallizes in the inverse spinel structure ($a_\mathrm{Fe_3O_4}=8.396\,$\r{A}) and exhibits mixed valences of $\mathrm{Fe^{2+}}$ and $\mathrm{Fe^{3+}}$ cations \cite{Coey}. W\"ustite (Fe\textsubscript{1-$\delta$}O) -- a semiconducting antiferromagnet with a N\'eel temperature of $T_\mathrm{N}=-75^\circ$C, crystallizing in the rock salt structure ($a_\mathrm{FeO}=4.332\,$\r{A}) \cite{Koch69,Battle79} -- represents the lowest oxidized polymorph. W\"ustite is often found in a defect stoichiometry and typically denoted as Fe\textsubscript{1-$\delta$}O, with $\delta$ ranging from 0.05 to 0.17 \cite{Koch69,Parkinson16}.
\begin{table*}[t]
\centering
\caption{\label{tab_p2lattice} Thicknesses and deposition rates of the samples.}
\begin{tabular}{c|ccc|cc}
~ & ~ & tr-HEXRD & ~  & tr-HAXPES & ~ \\ 
sample & $\mathrm{Fe_3O_4/NiO/MgO}$ & $\mathrm{Fe_3O_4/MgO}$ & $\mathrm{Fe_3O_4/SrTiO_3}$ & continuous & step-wise \\ 
\hline
%$a$ (\r{A}) & $8.23\pm 0.36$ & $8.46 \pm 0.35$ & $8.42 \pm 0.30$ \\ 
%$c$ (\r{A}) & $8.44\pm 0.24$ & $8.52 \pm 0.35$ & $8.40 \pm 0.32$ \\
$d_\mathrm{Fe_3O_4}$ (nm) & $18.9\pm 0.1$ & $8.6 \pm 0.2$  & $12.4\pm 0.5$ & $18.8\pm 0.3$ & $17.1\pm 0.7$ \\
$\mathrm{rate}$ (nm/min) & 2.6 & 0.86 & 0.31 & 0.47 & 0.86 \\
\end{tabular} 
\end{table*}

Because of this variability of their properties, thin films of iron oxides are often regarded as attractive for spintronics. For instance, magnetite is a long-standing candidate to contribute to all-oxide thin-film spintronic devices, as a source for spin-polarized currents \cite{Moussy13,Coey03,Bibes07,Zutic04,Moyer15,Marnitz15}, maghemite has been discussed as a magnetic tunnel barrier for spin-filter \cite{Yanagihara06,Grau-Crespo10}, and exchange bias has been observed in Fe/Fe\textsubscript{1-$\delta$}O bilayers.

In order to effectively study thin-film-based spintronic devices, a detailed knowledge of the growth mechanism down to the monolayer level is necessary. The growth mechanism of $\mathrm{Fe_3O_4}$ thin films has been frequently studied. Chang et al. investigated the cation stoichiometry of $\mathrm{Fe_3O_4/MgO(001)}$ for very thin films of few monolayers with x-ray absorption spectroscopy (XAS) and concluded that these ultrathin films dynamically redistribute during growth in order to avoid polarity \cite{Chang16}. 
It has also been reported on an iron-deficient w\"ustite layer of about 3 monolayers at the $\mathrm{Fe_3O_4/MgO(001)}$ interface observed by x-ray diffraction (XRD) \cite{Bertram12}.

$\mathrm{Fe_3O_4}$ ultrathin films have been grown on a wide array of substrates. The most widespread choice is the rock salt crystal MgO(001), because doubling its lattice constant of $a_\mathrm{MgO}=4.212\,$\r{A} results in a small mismatch of 0.3\% to $\mathrm{Fe_3O_4}$ and grants pseudomorphic growth \cite{Celotto03,Zaag00,Chang16,Bertram11,Bertram12,Bertram13,Margulies97,Tobin07,Arora08}.
A drawback of MgO(001) substrates is the limitation of growth and annealing temperatures to $250^\circ$C, as Mg starts to interdiffuse into the magnetite film at higher temperatures \cite{Kim09}. 
$\mathrm{SrTiO_3(001)}$ substrates, in contrast, crystallize in the perovskite structure and have a lattice constant of $a_\mathrm{SrTiO_3}=3.905\,$\r{A} and a  mismatch of $-7.5$\% to magnetite, offering the possibility to study strain effects on magnetite \cite{Monti13,Kale01,Rubio15,Kuschel16,Kuschel17}. Different than on MgO, $\mathrm{Fe_3O_4}$ can be grown on $\mathrm{SrTiO_3}$ at elevated temperatures with no risk of interdiffusion \cite{Kuschel16}. On SrTiO\textsubscript{3}(001), $\mathrm{Fe_3O_4}$  has been reported to grow in different orientations: for temperatures below $400^\circ$, it grows with a (001) orientation, while for temperatures of $700^\circ$ and above, the more stable (111) orientation is favored \cite{Takahashi12, Takahashi14}.
%, and it may be doped by Niobium in order to tune its conductivity \cite{Kuepper16}. 
Ultrathin film bilayers of $\mathrm{Fe_3O_4}$ and $\mathrm{NiO}$, on the other hand, are very interesting from a spintronic perspective, because the exchange bias between the ferrimagnetic magnetite and NiO, which is antiferromagnetic below its N\'eel temprature of $T_\mathrm{N}=250^\circ$C, can be exploited for magnetic tunnel junctions  \cite{Keller02,Gatel05,Krug08,Pilard07,Kuepper16}. This effect causes an asymmetric hysteresis of the ferromagnetic film, with different switching fields depending on the direction of the external magnetic field. 

In previous studies, $\mathrm{Fe_3O_4}$ films were grown by reactive molecular beam epitaxy (RMBE) and subsequently investigated after growth \cite{Schemme15,Kuepper16}. XRD has been used to study thickness dependent structural properties while x-ray photoelectron spectroscopy (XPS) has been applied for electronic and chemical analysis. A more direct access to the growth process can be granted by simultaneously depositing the film and performing XRD \cite{Kuschel17} and XPS investigations. This is the route taken in this study. 
$\mathrm{Fe_3O_4}$ ultrathin films are grown on Nb-doped $\mathrm{SrTiO_3(001)}$, $\mathrm{MgO(001)}$ and $\mathrm{NiO/MgO(001)}$ by RMBE, and time-resolved high-energy x-ray diffraction (tr-HEXRD) is used to observe the formation of Bragg peaks of the evolving $\mathrm{Fe_3O_4}$ film which are specific to the order of octahedrally and tetrahedrally coordinated Fe cations during growth. Along with the tr-HEXRD, time-resolved grazing incidence small angle x-ray scattering (tr-GISAXS) data are recorded in order to monitor the growth mode of the films.
Time-resolved hard x-ray photoelectron spectroscopy (tr-HAXPES) is employed to observe the development of the $\mathrm{Fe~2p}$ spectrum of $\mathrm{Fe_3O_4/SrTiO_3}$ during the deposition in order to allow conclusions towards the different oxidation states of Fe through the entire film thickness.

\section{\label{sec:level2}Experimental details}
Synchrotron-based tr-HEXRD measurements utilize high energy x-rays in combination with large area 2D detectors to collect time-resolved diffraction data of dynamic processes, such as thin film growth \cite{Roelsgaard19} or catalytic processes \cite{Gustafson14}. %The aim of this study is to distinguish between the octahedral order, present in both $\mathrm{Fe_{1-\delta}O}$ and $\mathrm{Fe_3O_4}$, and the tetrahedral order exclusive to $\mathrm{Fe_3O_4}$. The inverse spinel structure of $\mathrm{Fe_3O_4}$ can be described as consisting of a cubic close-packed oxygen sublattice, a sublattice with double periodicity containing the octahedrally coordinated $\mathrm{Fe^{2+,3+}_{oct}}$ cations (B-sites) and a sublattice containing the tetrahedrally coordinated $\mathrm{Fe^{3+}_{tet}}$ cations (A-sites). These sublattices give rise to different Bragg-reflections, some of which are exclusive to the spinel structure while others also occur in the $\mathrm{Fe_{1-\delta}O}$ rock salt phase \cite{Bertram13}.
In this study, we observe the intensity evolution of the Bragg reflections of $\mathrm{Fe_3O_4}$ during the deposition of the films. 

Our tr-HEXRD and tr-GISAXS measurements were performed at beamline P07/EH2 of PETRA III at Deutsches Elektronen-Synchrotron (DESY). A custom-designed UHV-deposition chamber was mounted on the diffractometer, in order to perform grazing-incidence diffraction with a glancing angle of $\theta = 0.03^\circ$ during the deposition of the thin films.
The sample preparation followed the procedures in Refs. \cite{Kuepper16,Schemme15,Rodewald19}.
 Before deposition, the MgO and the Nb-doped ($0.05\,$ wt\%) $\mathrm{SrTiO_3}$ substrates were annealed at $400^\circ$C for one hour in an oxygen atmosphere of $p_\mathrm{O_2}=10^{-4}\,$mbar. 
For the $\mathrm{Fe_3O_4/NiO/MgO(001)}$ sample, first a $5.6\,$nm thick NiO layer was grown by evaporating Ni from a metal target in an oxygen atmosphere of $p_\mathrm{O_2}=5\times 10^{-6}\,$mbar at $250^\circ$C substrate temperature. Both the $\mathrm{Fe_3O_4}$ films on $\mathrm{NiO/MgO(001)}$ and on $\mathrm{MgO(001)}$ were deposited by evaporating Fe under the same conditions as the NiO film. The $\mathrm{Fe_3O_4/SrTiO_3(001)}$ film was deposited in a reduced oxygen atmosphere of $p_\mathrm{O_2}=1\times 10^{-6}\,$mbar and a higher substrate temperature of $350^\circ$C. 
Film thicknesses were controlled by calibrating the fluxes of the evaporators. Table \ref{tab_p2lattice} summarizes the final thicknesses $d_\mathrm{Fe_3O_4}$ after growth and growth rates for the three investigated samples.
%\begin{figure}[t]
%\centering
%\includegraphics[scale=0.4]{Bilder/Fe3O4_RSM_v3.pdf}
%\caption{Reciprocal space map of the $\mathrm{Fe_3O_4/MgO(001)}$ sample. The right half shows the measurement, the left half a schematic of the Bragg positions for MgO (black circles), $\mathrm{SrTiO_3}$ (green crosses) and $\mathrm{Fe_3O_4}$. Open red squares denote reflections that occur exclusively for the spinel structure, while reflections marked by blue squares are also allowed for the rock salt structure. The white dashed box marks the region of the (222) and the (224) reflections whose time evolution during growth is monitored in Figs. \ref{fig_p2sto} and  \ref{fig_p2MgO}.   
%The white disk at the (222) position is a semi-transparent beam stop protecting the detector from the bright substrate (111) reflection, and the white grid stem from module borders of the detector. The Bragg reflections labeled in the schematic are in $\mathrm{Fe_3O_4}$ lattice units.}
%\label{fig_p2RSM}
%\end{figure}

\begin{figure}
\centering
\includegraphics[scale=0.4]{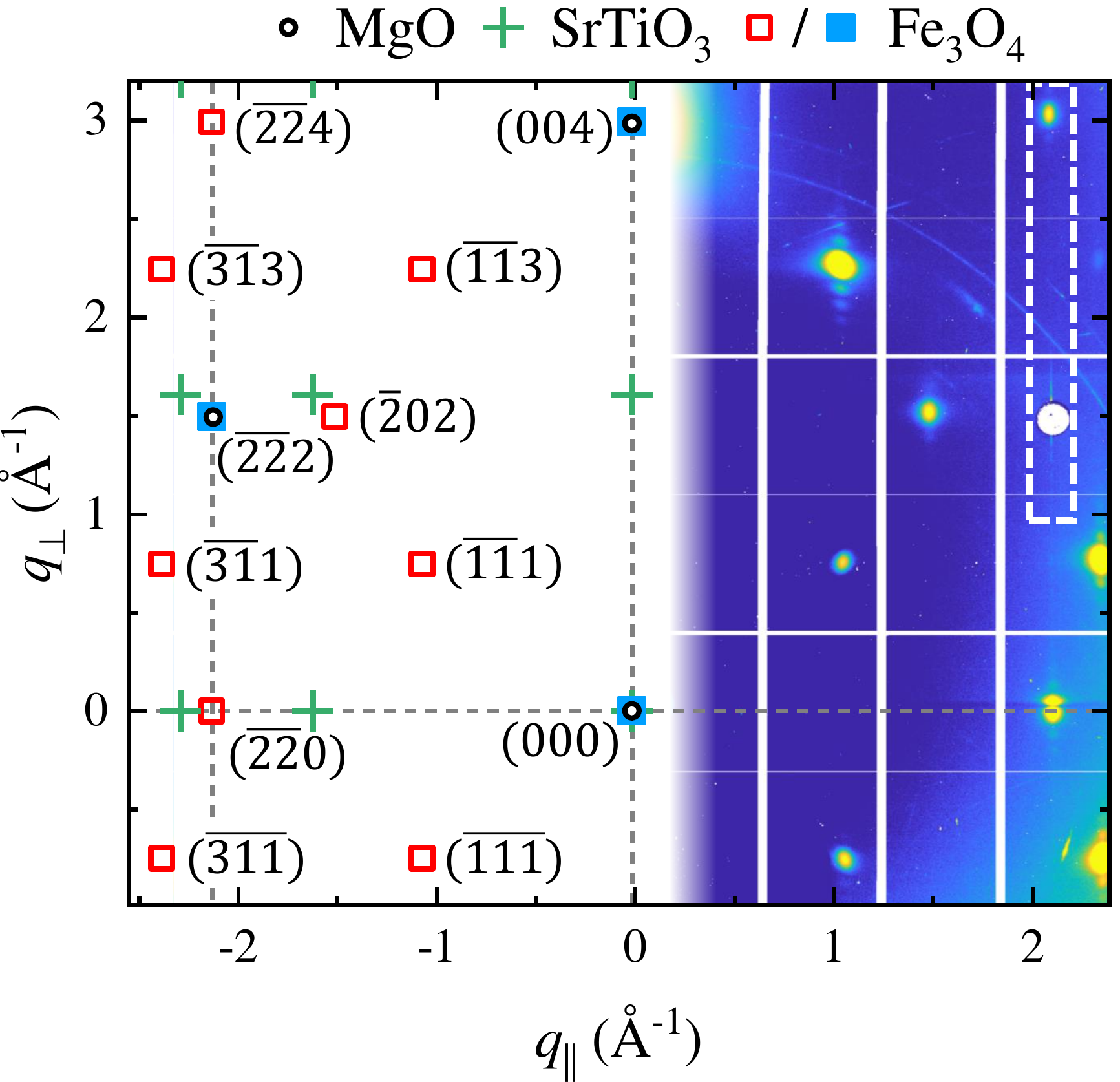}
\caption{Reciprocal space map of the $\mathrm{Fe_3O_4}$/MgO(001) sample. The right half shows the measurement, the left half a schematic of the Bragg positions for MgO (black circles), SrTiO\textsubscript{3} (green crosses) and $\mathrm{Fe_3O_4}$. Open red squares denote $\mathrm{Fe_3O_4}$  reflections that occur exclusively for the spinel structure, while reflections marked by blue squares are also allowed for the rock salt structure. The white dashed box marks the region of the (222) and the (224) reflections whose time evolution during growth is monitored in Figs. \ref{fig_p2sto} and  \ref{fig_p2MgO}.   
The white disk at the (222) position is a semi-transparent beam stop protecting the detector from the bright substrate (111) reflection, and the white grid stem from module borders of the detector. The Bragg reflections labeled in the schematic are in $\mathrm{Fe_3O_4}$  lattice units.
\label{fig_p2RSM}
}
\end{figure}

For the $\mathrm{Fe_3O_4/MgO(001)}$ and the $\mathrm{Fe_3O_4/SrTiO_3(001)}$ samples, a photon energy of $74\,$keV was used, and data were recorded on a Dectris Pilatus 3X CdTe 2D area detector. Before the start of the deposition, the samples were azimuthally aligned to an angle $\omega_{(111)}$ at which the Bragg condition for the substrate (111) reflection was fulfilled.
During deposition, the samples were continuously azimuthally rotated between $\omega_{(111)} \pm 7^\circ$ with a rotation speed of $\mathrm{2^\circ/sec}$, in order to observe the development of the $(22L)_\mathrm{Fe_3O_4}$ crystal truncation rod (CTR) during growth. The detector images obtained during each one of these $14^\circ$-rotations were then summed up to obtain one reciprocal space map (RSM) every 12 seconds (cf. right half of Fig.~\ref{fig_p2RSM}).
After growth, the electronic structure of the samples was characterized \textit{in situ} by XPS, using a Phoibos HSA 150 hemispherical analyzer and an Al K$\alpha$ anode, in order to probe the  stoichiometry of the grown iron oxide, and a RSM with a full rotation of $90^\circ$ was recorded with a rotation speed of $0.5^\circ/\mathrm{sec}$.

For the $\mathrm{Fe_3O_4/NiO/MgO(001)}$ sample, we followed the same procedure, but using a photon energy of $72\,$keV and a Perkin-Elmer XRD1621 detector, with a rotation range of $\omega_{(111)} \pm 5^\circ$, and recorded one RSM every 28 seconds. Again, after growth a RSM with a full rotation of $90^\circ$ was recorded.

In order to obtain information on the temporal evolution of the stoichiometry of the different cations of $\mathrm{Fe_3O_4}$  during the growth, we performed complimentary tr-HAXPES measurements during the growth of $\mathrm{Fe_3O_4}$ films on $\mathrm{SrTiO_3}$. To this end, we installed the same custom-designed UHV-deposition chamber used for the tr-HEXRD measurements at beamline P22 of PETRA III at DESY \cite{P22}. The samples were illuminated by an x-ray beam under a glancing angle of $3^\circ$ at a photon energy of $4.6\,$keV, and photoelectrons were collected using the Phoibos HSA 150 hemispherical analyzer at a $\alpha = 30^\circ$ angle from the surface normal, resulting in an information depth of about $\mathrm{ID(95)}=16\,$nm \cite{Kuepper16}. 

The films were deposited in an oxygen atmosphere of $p_\mathrm{O_2}=5\times 10^{-6}\,$mbar at a substrate temperature of $400^\circ$C.
One sample was deposited continuously up to a thickness of $18.8\,$nm, and one sample was deposited step-wise:
After each spectrum taken during the deposition of $1.7\,$nm $\mathrm{Fe_3O_4}$, the growth was stopped by interrupting the molecular beam, and a $\mathrm{Fe~2p}$ spectrum was measured with higher sampling time before the next deposition step.
The photoelectron spectra of the $\mathrm{Fe~2p}$ were recorded by scanning from high to low binding energies. 
For the continuously deposited film, the spectra were recorded in $0.2\,$eV energy steps and an integration time of $0.1\,\text{sec/step}$, taking 60 seconds for a single spectrum, which means that the film thickness was increasing by $0.5\,$nm during each measurement.
For instance, the first spectrum started recording at a binding energy of $760\,$eV and a film thickness of $0\,$nm, and ended at a binding energy of $690\,$eV with a film thickness of $0.5\,$nm. We used a linear scaling function on these spectra in order to account for the fact that the intensity of the $\mathrm{Fe~2p}$ increases with the film thickness. 
The step-wise deposited film in energy steps of $0.2\,$eV and an integration time of $0.2\,\text{sec/step}$, taking 120 seconds for a single spectrum. The deposition rates can be found in Tab. \ref{tab_p2lattice}.
Charge transfer multiplet (CTM) calculations of the XPS spectra of the three cation species in $\mathrm{Fe_3O_4}$  have been performed using the method and parameter set of Fuji et al.  \cite{Fuji99}.

\section{\label{sec:level4}Results}
\subsection{Reciprocal space map of Fe\textsubscript{3}O\textsubscript{4}}
Figure \ref{fig_p2RSM} shows a RSM of the as-grown $\mathrm{Fe_3O_4/MgO}$ sample, obtained by a full $90^\circ$ azimuthal rotation. The right half shows recorded data and the left a schematic of the peak positions from MgO, $\mathrm{SrTiO_3}$ and $\mathrm{Fe_3O_4}$. %From the positions of the spinel reflections compared to the substrate reflections, we can derive the in-plane and out-of-plane lattice constants $a$ and $c$ of the $\mathrm{Fe_3O_4}$, given in Tab. \ref{tab_p2lattice}.

\begin{figure}
\centering
\includegraphics[scale=0.4]{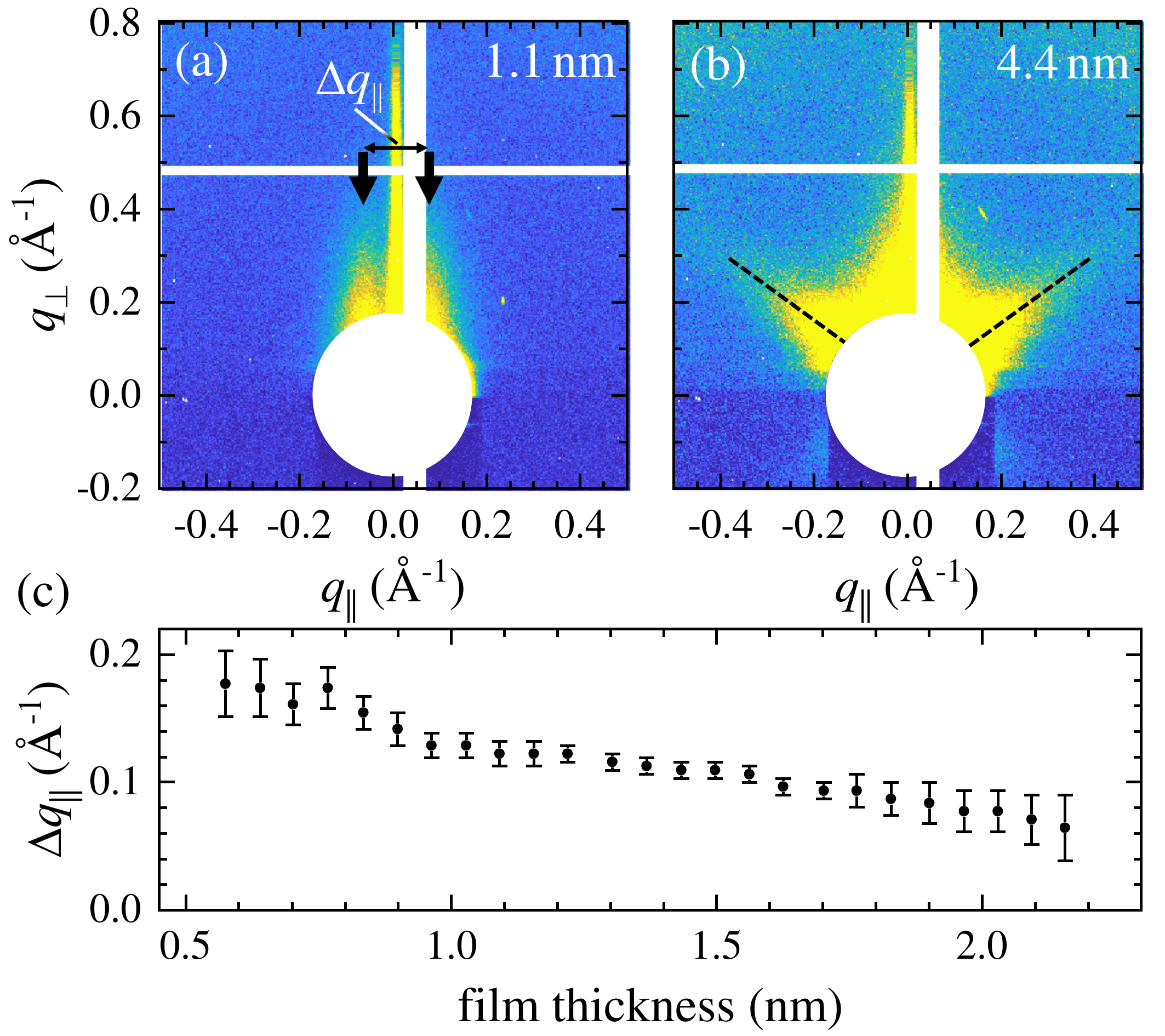}
\caption{(a) Small angle region of an RSM taken at a thickness of 1.1$\,$nm of the $\mathrm{Fe_3O_4}$/SrTiO\textsubscript{3} film. Black arrows indicate two distinct streaks parallel to \textit{q}\textsubscript{$\perp$}. (b) The same region as in (a), but at a film thickness of 4.4$\,$nm. Two streaks at an angle of 55$^\circ$ with respect to the surface normal are highlighted by dashed black lines. (c) Reciprocal space distance $\Delta q_\parallel$ of the two streaks in (a) as a function of film thickness.}
\label{fig_p2SAXS}
\end{figure}

The goal of the tr-HEXRD measurements is to record the intensity evolution of the $(222)_\mathrm{Fe_3O_4}$ and the $(224)_\mathrm{Fe_3O_4}$ reflection. This endeavor is complicated by the close lattice match of MgO and $\mathrm{Fe_3O_4}$:
$\mathrm{Fe_3O_4}$ has roughly double the lattice constant of MgO with a small lattice mismatch of only 0.3\%, and consequently, Bragg reflections $(HKL)_\mathrm{MgO}$ of MgO almost coincide with reflections $(2 H,2 K,2 L)_\mathrm{Fe_3O_4}$ of $\mathrm{Fe_3O_4}$. 
This can be seen in the schematic RSM in Fig.~\ref{fig_p2RSM}, where black circles indicate the MgO reflections, and squares the reflections of $\mathrm{Fe_3O_4}$.
The intensity difference between the bright substrate reflections and the weaker film reflections is too high for the detector to record them simultaneously. For this reason, the MgO reflections had to be covered by semi-transparent beamstops on the detector. For instance, on the one hand, the $(222)_\mathrm{Fe_3O_4}$ has to be blocked (see white disk in the RSM presented in the right half of Fig.~\ref{fig_p2RSM}) since it almost coincides with the (111)-reflection of MgO. On the other hand, the $(224)_\mathrm{Fe_3O_4}$ is fully visible, since the corresponding substrate-related Bragg reflection $(112)_\mathrm{MgO}$ is forbidden. 

In contrast, the perovskite structure of $\mathrm{SrTiO_3}$ has more allowed reflections, including the $(112)_\mathrm{SrTiO_3}$, as illustrated by the green crosses in Fig.~\ref{fig_p2RSM}. But since its higher lattice mismatch of $-7.5$\% to $\mathrm{Fe_3O_4}$, the substrate reflections are well separated from the film reflection, making a direct observation of both the $(222)_\mathrm{Fe_3O_4}$ and the $(224)_\mathrm{Fe_3O_4}$ reflections possible.

\subsection{Fe\textsubscript{3}O\textsubscript{4}/SrTiO\textsubscript{3}}
\begin{figure}[t]
\centering
\includegraphics[scale=0.4]{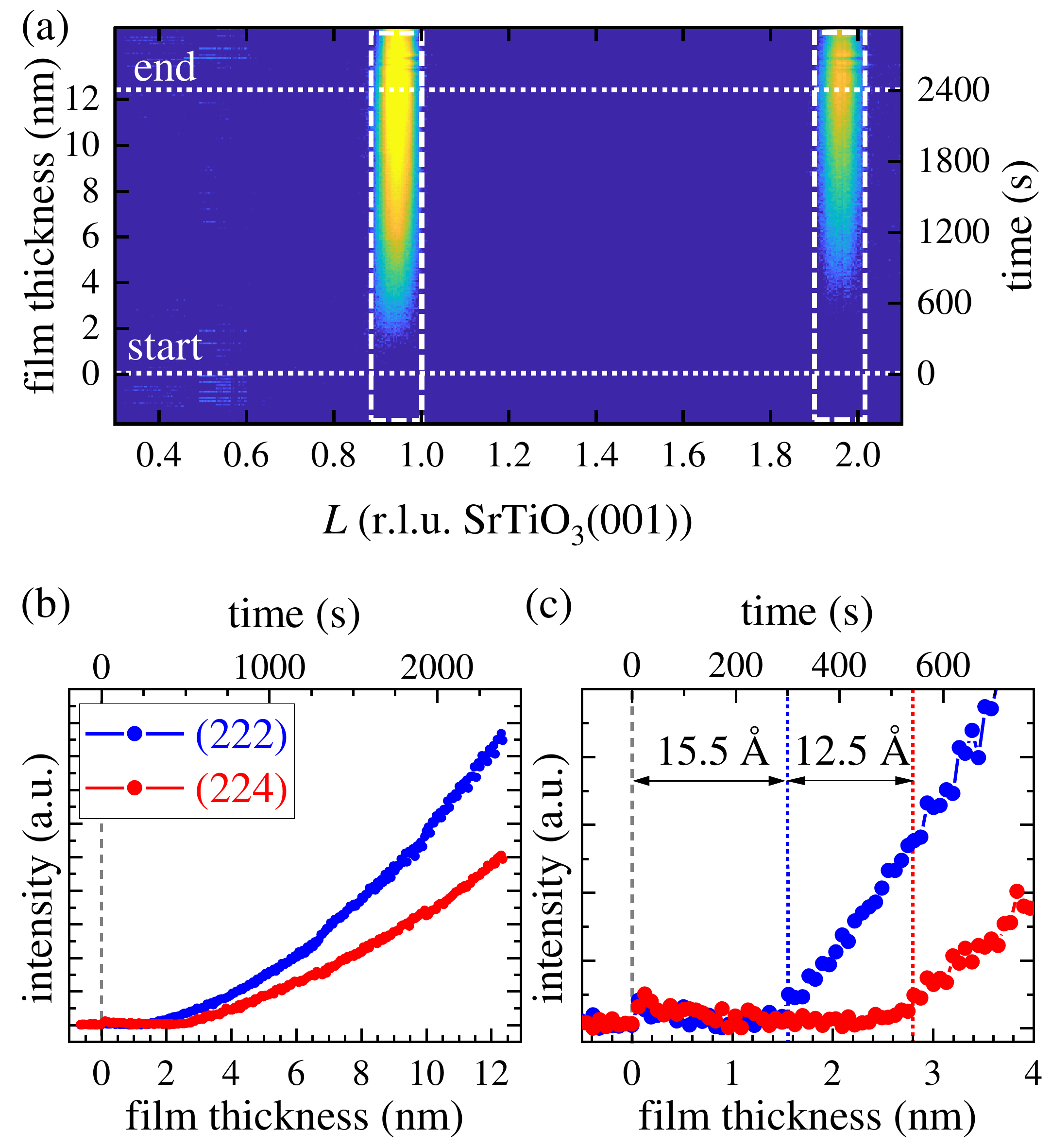}
\caption{%(a) Contributions of the octahedral, tetrahedral and oxygen sites to the structure factor $|\mathrm{F_{site}}|^2$ of $\mathrm{Fe_3O_4}$. 
(a) Integrated diffracted intensity along the (22\textit{L})\textsubscript{$\mathrm{Fe_3O_4}$}-rod, in the region indicated by the white box in Fig.~\ref{fig_p2RSM}, as a function of film thickness. The dotted horizontal white lines indicate the start and end of the deposition process. The white dashed boxes show the regions where the intensities for the (222)\textsubscript{$\mathrm{Fe_3O_4}$} and (224)\textsubscript{$\mathrm{Fe_3O_4}$  } presented in (b),(c) are taken, respectively. (b) Intensity evolution for the (222)\textsubscript{$\mathrm{Fe_3O_4}$} and (224)\textsubscript{$\mathrm{Fe_3O_4}$} during deposition. (c) Close-up of the initial growth stage. Blue and red dotted lines indicate film thicknesses where the (222)\textsubscript{$\mathrm{Fe_3O_4}$} and (224)\textsubscript{$\mathrm{Fe_3O_4}$} start to change intensity.}
\label{fig_p2sto}
\end{figure}
Information on the early growth stage can be collected by tr-HEXRD by analyzing the grazing incidence small angle x-ray scattering (GISAXS) data in the small angle region of the RSMs.
Figure \ref{fig_p2SAXS}(a) shows the small angle region of an RSM taken at a film thickness of $1.1\,$nm. Two streaks parallel to the $q_\perp$ direction are highlighted by two black arrows. Their reciprocal space distance $\Delta q_\parallel$ is plotted as a function of film thickness in Fig. \ref{fig_p2SAXS}(c). They become first visible at a film thickness of about $0.5\,$nm, and then move gradually closer to each other along the $q_\parallel$ direction with increasing film thickness, until they eventually become indistinguishable and merge in the center at a thickness of about $2.2\,$nm. 
At a film thickness of $2.9\,$nm, two new streaks appear, highlighted by black dashed lines in Fig.~\ref{fig_p2SAXS}(b). In the projection of the RSM, they are tilted by an angle of $55^\circ$ from the $q_\perp$ direction. Such streaks are characteristic for (111) facets \cite{Renaud09}
and remain present for the rest of the growth process, including the finished film.

\begin{figure}[t]
\centering
\includegraphics[scale=0.4]{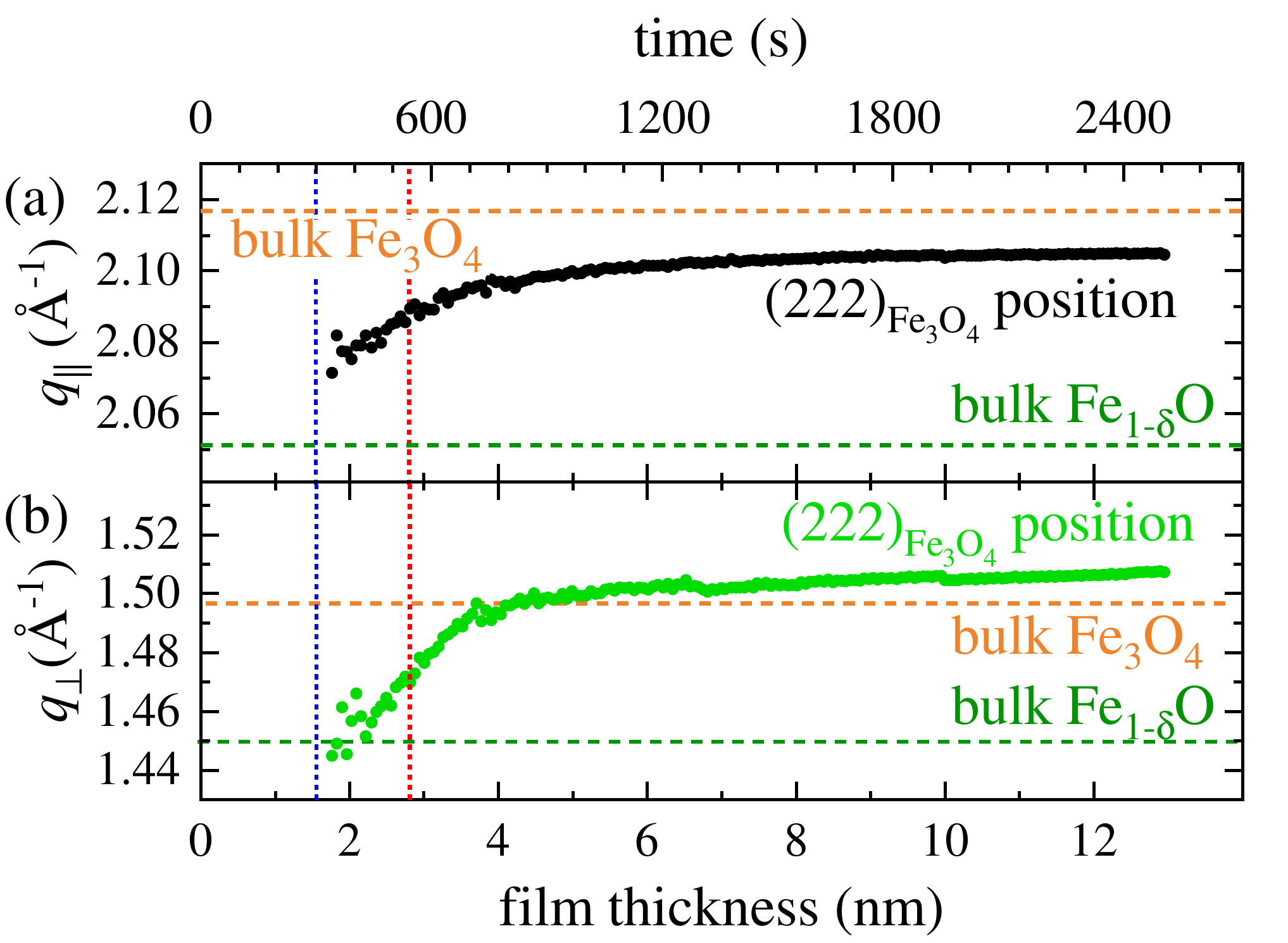}
\caption{Evolution of (a) the in-plane scattering vector \textit{q\textsubscript{$\perp$}} and (b) the out-of-plane scattering vector \textit{q\textsubscript{$\parallel$}} of the film (222) reflection. Dashed horizontal lines indicate the expected scattering vectors for bulk $\mathrm{Fe_3O_4}$  and bulk Fe\textsubscript{1-$\delta$}O. Dotted blue and red vertical lines indicate the thicknesses at which the (222) and the (224) reflections emerge, respectively.}
\label{fig_p2lats}
\end{figure}

\begin{figure*}[t]
\centering
\includegraphics[scale=0.4]{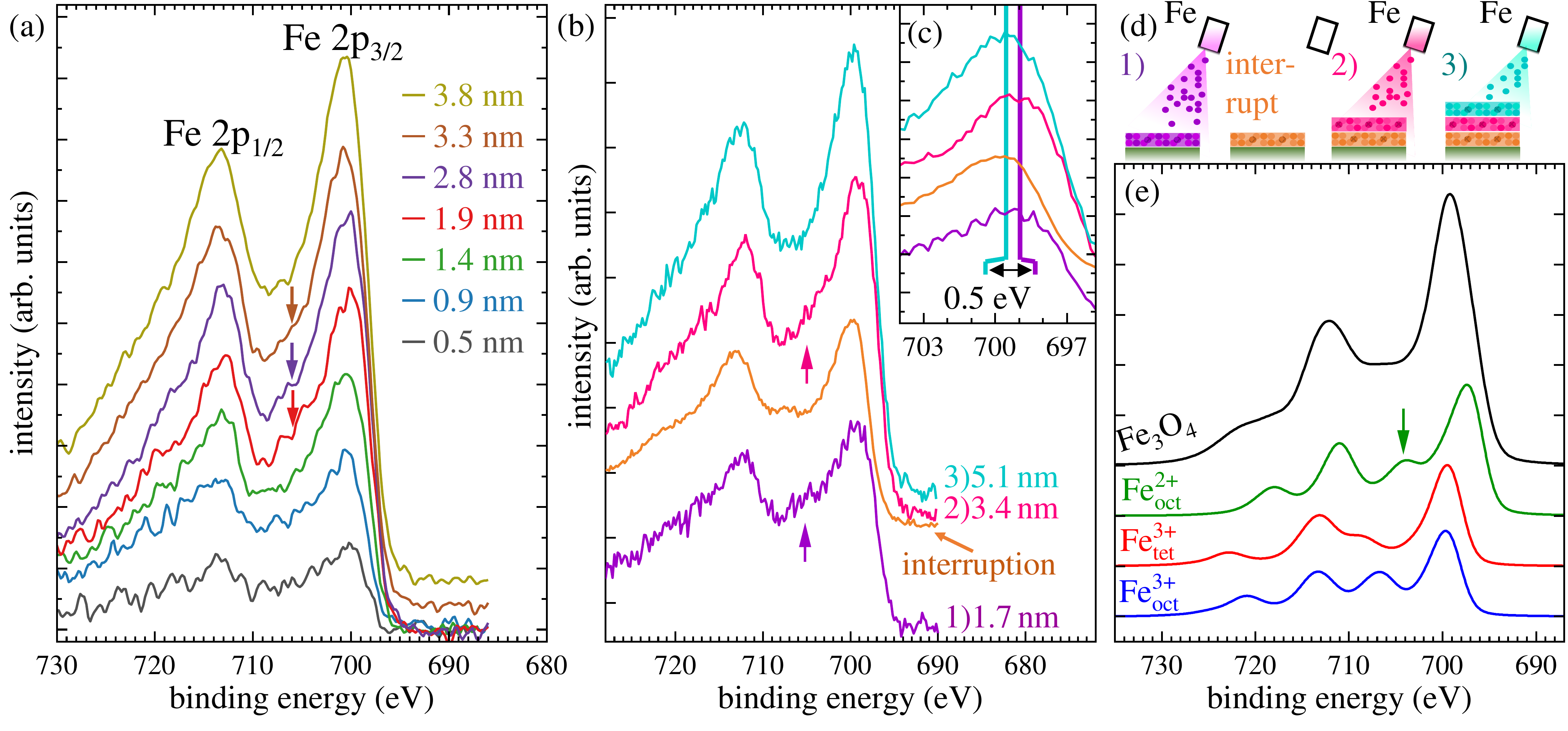}
\caption{(a) HAXPES data of the Fe 2p recorded during the continuous deposition of a $\mathrm{Fe_3O_4}$/SrTiO\textsubscript{3} film. The data was smoothed by a Gaussian filter. (b) HAXPES data of the Fe 2p recorded during the step-wise deposition process of a $\mathrm{Fe_3O_4}$/SrTiO\textsubscript{3} film [1) purple, 2) magenta and 3) cyan] and during an interruption of the deposition (orange), as illustrated in (d). At the end of the measurements during deposition, the films were 1.7$\,$nm, 3.4$\,$nm and 5.1$\,$nm thick. (c) Close-up of the Fe 2p\textsubscript{3/2} region; violet and yellow lines indicate center of the peak in the respective spectrum. (d) Sketch of the step-wise deposition process, corresponding to the spectra in (b). (e) CTM calculations of the XPS spectra for the three cation species of $\mathrm{Fe_3O_4}$, $\mathrm{Fe^{2+}_{oct}}$ (green), $\mathrm{Fe^{3+}_{tet}}$ (red) and $\mathrm{Fe^{3+}_{oct}}$ (blue). Grey line is the sum of the $\mathrm{Fe^{2+}_{oct}}$ and $\mathrm{Fe^{3+}_{oct}}$ spectra, and the black line is the sum of all three spectra, representing the full $\mathrm{Fe_3O_4}$ spectrum.
Arrows in (a),(b) and (e) highlight the charge-transfer satellite characteristic for Fe\textsuperscript{2+} cations.
}
\label{fig_p2HAX}
\end{figure*}
In order to gain insight into the growth dynamics of the $\mathrm{Fe_3O_4/SrTiO_3}$ film, we used tr-HEXRD to monitor the evolution of the $(222)_\mathrm{Fe_3O_4}$ and the $(224)_\mathrm{Fe_3O_4}$ reflections, which are sensitive to the octahedral and tetrahedral order, respectively. 
Figure~\ref{fig_p2sto}(a) shows how the intensity along the $(22L)_\mathrm{Fe_3O_4}$-rod (white dashed box in Fig.~\ref{fig_p2RSM}) changes with the film thickness. The color scale corresponds to the intensity, the horizontal axis to the Miller index $L$ (parallel to $q_\perp$) and the vertical axis to the film thickness.
For the thickness axis, we assumed a linear relation between deposition time and film thickness. In the color plot of Fig.~\ref{fig_p2sto}(a), it can be seen that at the beginning of the growth, neither the $(222)_\mathrm{Fe_3O_4}$ nor the $(224)_\mathrm{Fe_3O_4}$ can be observed. After about $1\,$nm of growth, intensity is detected at the $(222)_\mathrm{Fe_3O_4}$ position, and with a slight delay the $(224)_\mathrm{Fe_3O_4}$ reflection appears. 
For better quantification, Fig.~\ref{fig_p2sto}(b) shows the $L$-integrated intensity along the white dashed boxes for the $(222)_\mathrm{Fe_3O_4}$ and the $(224)_\mathrm{Fe_3O_4}$ (cf. Fig.~\ref{fig_p2sto}(a)) as a function of film thickness and deposition time for the full growth process, and Fig.~\ref{fig_p2sto}(c) contains a close-up of the early growth stage. Here, the delay between the emergence of the $(222)_\mathrm{Fe_3O_4}$ and the $(224)_\mathrm{Fe_3O_4}$ can be determined to be $\Delta d=1.25\pm 0.12\,$nm. It can also be seen that the $(222)_\mathrm{Fe_3O_4}$ does not appear before a coverage of $1.55\,$nm has been reached.

The position of the $(222)_\mathrm{Fe_3O_4}$ has been fitted to draw conclusions towards the lattice constant during growth. Figures \ref{fig_p2lats}(a),(b) show the evolution of the in-plane component $q_\parallel$ and the out-of-plane component $q_\perp$ of the scattering vector of the $(222)_\mathrm{Fe_3O_4}$. Both components move from a smaller position at low film thicknesses to a larger position at thicker films. This corresponds to a progressive compression of the lattice parameters in both vertical and lateral direction for increasing film thickness.
The dashed lines indicate the scattering vector components expected for bulk $\mathrm{Fe_3O_4}$  and for bulk Fe\textsubscript{1-$\delta$}O. It can be seen that the position of the (222) reflection is close to the expected value for Fe\textsubscript{1-$\delta$}O at the beginning of the deposition, and settles close to the position expected for $\mathrm{Fe_3O_4}$  when the film grows thicker.
\begin{figure*}[t]
\centering
\includegraphics[scale=0.4]{{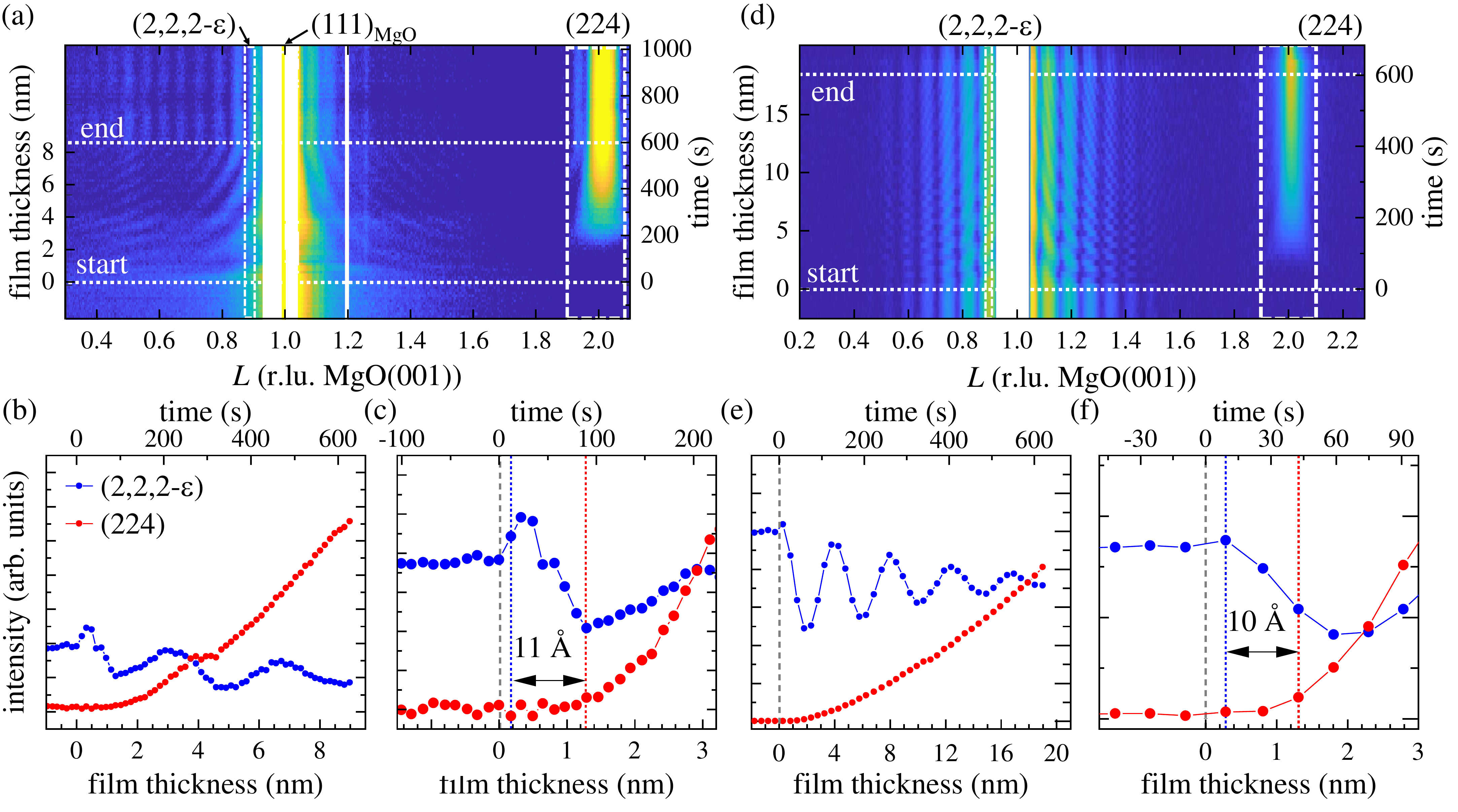}}
\caption{(a),(d) Diffracted intensity along the (22L)\textsubscript{$\mathrm{Fe_3O_4}$ }-CTR of the (a) $\mathrm{Fe_3O_4}$  film on MgO and the (d) $\mathrm{Fe_3O_4}$ film on MgO with NiO interlayer, in the region indicated by the white box in Fig.~\ref{fig_p2RSM}, as a function of film thickness. The dotted white lines indicate the start and end of the deposition process. The white area around \textit{L}=1 is caused by the semi-transparent beam stop for the (111)\textsubscript{MgO} substrate reflection, and the white line at L=1.2 stems from a module border of the detector. The white dashed boxes show the regions in which the intensities for the \mbox{(2,2,2-$\epsilon$)\textsubscript{$\mathrm{Fe_3O_4}$}} and (224)\textsubscript{$\mathrm{Fe_3O_4}$} in (b),(e) are taken. (b),(e) Intensity evolution for the (2,2,2-$\epsilon$)\textsubscript{Fe\textsubscript{3}O\textsubscript{4}} and (224)\textsubscript{$\mathrm{Fe_3O_4}$ } during the deposition of (b) the $\mathrm{Fe_3O_4}$/MgO film and (e) the $\mathrm{Fe_3O_4}$/NiO/MgO. (c),(f) Close-up of the initial growth stages in (b),(e) demonstrating the delayed onset of occupation of the tetrahedral sublattice. Data for the (2,2,2-$\epsilon$)\textsubscript{$\mathrm{Fe_3O_4}$} in (f) have been offset for better visibility. Blue and red dotted lines indicate the smallest film thickness where intensity is observed for the (2,2,2-$\epsilon$)\textsubscript{$\mathrm{Fe_3O_4}$} and (224)\textsubscript{$\mathrm{Fe_3O_4}$} reflections, respectively.}
\label{fig_p2MgO}
\end{figure*}

In order to make further conclusions about the oxide phase in this early growth phase, we performed tr-HAXPES measurements of the growth of $\mathrm{Fe_3O_4/SrTiO_3}$ films. The results are shown in Fig.~\ref{fig_p2HAX}.
Figure~\ref{fig_p2HAX}(a) shows the spectra recorded from the continuously deposited $\mathrm{Fe_3O_4}$/SrTiO\textsubscript{3} film. During the measurement of each spectrum, about $0.47\,$nm of $\mathrm{Fe_3O_4}$  have been deposited. For the films below $1.4\,$nm thickness, the low signal-to-noise ratio makes it difficult to observe spectral features. However, for film thicknesses from $1.9\,$nm to $3.3\,$nm, a satellite feature at the high-energy side of the Fe~$\mathrm{2p_{3/2}}$ can be observed, highlighted by the arrows. For films thicker than $3.8\,$nm, this satellite disappears, and the region between the Fe~$\mathrm{2p_{3/2}}$ and the Fe~$\mathrm{2p_{1/2}}$ is flat.
This effect is even more obvious in a step-wise deposition process, for which spectra could be recorded at higher integration times (cf. Fig.~\ref{fig_p2HAX}(b)). It was grown in steps of $1.7\,$nm. After each deposition step, the growth was interrupted. Spectra were taken during and after each deposition step.
The violet, magenta and cyan spectra presented in Fig.~\ref{fig_p2HAX}(b) were recorded during the first three deposition steps, and had thicknesses of $1.7\,$nm, $3.4\,$nm and $5.1\,$nm after the end of the corresponding measurement, respectively. The orange spectrum was taken during the interruption after the first deposition step, at a film thickness of $1.7\,$nm. These deposition steps are illustrated in Fig.~\ref{fig_p2HAX}(d), with the step labels 1), 2) and 3) corresponding to the spectra in Fig.~\ref{fig_p2HAX}(b).

For the spectra recorded during the first two deposition steps (purple and magenta), a satellite can be seen on the high-energy side of the $\mathrm{Fe~2p_{3/2}}$ (purple and magenta arrows), which vanishes for both, films thicker than $5.1\,$nm and the spectrum taken during the interruption. This satellite is well known to be a charge-transfer satellite stemming from $\mathrm{Fe^{2+}}$ cations on octahedral lattice sites. To illustrate this, Fig.~\ref{fig_p2HAX}(e) shows CTM calculations of the $\mathrm{Fe~2p}$ spectrum for the three cation species in $\mathrm{Fe_3O_4}$. The individual cation spectra show a distinct charge-transfer satellite between the $\mathrm{Fe~2p_{3/2}}$ and the $\mathrm{Fe~2p_{1/2}}$ lines at different energies. It is highlighted for the $\mathrm{Fe^{2+}_{oct}}$  spectrum by a green arrow. Summing up the three spectra in a 1:1:1 ratio yields the $\mathrm{Fe_3O_4}$ spectrum (black), which does not exhibit any apparent satellite structure between the $\mathrm{Fe~2p}$ main lines, because the satellites of the individual cation spectra overlap in such a way that they form a flat plateau \cite{Fuji99}. This shape is observed for the spectra during the third deposition step and in the interruption after the first deposition step in Fig.~\ref{fig_p2HAX}(b), and for film thicknesses higher than $3.8\,$nm in Fig.~\ref{fig_p2HAX}(a).

Additionally, we observe a chemical shift of about $0.5\,$eV for both $\mathrm{Fe~2p}$ main lines between the spectrum during the first deposition step on the one hand and the spectra recorded after the first and during the third step on the other hand, shown for the $\mathrm{Fe~2p_{3/2}}$ peak in the inset of Fig.~\ref{fig_p2HAX}(a). The spectrum taken during the second deposition step displays both a weaker satellite as well as an energy position in between the two other spectra.
%Both, the satellite and the chemical shift in the spectra for the two thinner films in Fig.~\ref{fig_p2HAX}(a) is an indicator for an excess of $\mathrm{Fe^{2+}}$ cations.

%This result may suggest a $\mathrm{Fe_{1-\delta}O}$ phase, but is not the only possibility. For example, the grey line in Fig.~ \ref{fig_p2HAX}(b) is the sum of the $\mathrm{Fe^{2+}_{oct}}$ and the $\mathrm{Fe^{3+}_{oct}}$ spectra, representing a $\mathrm{Fe^{3+}_{tet}}$-depleted magnetite phase. This simulation exhibits a very similar satellite, and shows a chemical shift of about $0.45\,$eV found in the data. 

\subsection{Fe\textsubscript{3}O\textsubscript{4}/MgO and Fe\textsubscript{3}O\textsubscript{4}/NiO/MgO}
For the $\mathrm{Fe_3O_4}$ films deposited on MgO, an according evaluation of the tr-HEXRD data is slightly more difficult because the overlap of the $(222)_\mathrm{Fe_3O_4}$ and the $(111)_\mathrm{MgO}$ reflections does not allow for an immediate observation of the $(222)_\mathrm{Fe_3O_4}$. Particularly, the region around the $(111)_\mathrm{MgO}$ has to be blocked by a semi-transparent absorber, as can be seen in Fig.~\ref{fig_p2RSM}.
Therefore, instead of the Bragg peak, we observe the intensity of the CTR at a position $(2,2,2-\epsilon)_\mathrm{Fe_3O_4}$ close to the $(222)_\mathrm{Fe_3O_4}$ reflection, with $\epsilon=0.12$.
Analogous to Fig.~\ref{fig_p2sto}(a), Fig.~\ref{fig_p2MgO}(a) shows a false color map of the diffracted intensity during the growth process of the $\mathrm{Fe_3O_4}$ film on MgO. Before starting the growth process, again, there is no intensity at the $(224)_\mathrm{Fe_3O_4}$ position. Around the position of the $(222)_\mathrm{Fe_3O_4}$ ($L \simeq 1.003$ in Fig.~\ref{fig_p2MgO}(a)), however, no intensity (white region) can be detected due to the semi-transparent absorber. At $L=1$, the $(111)_\mathrm{MgO}$ reflection is visible as a sharp bright line. The intensity has been corrected for the attenuation by the semi-transparent absorber. In the vicinity of the absorber, the $(111)_\mathrm{MgO}$-CTR decays monotonically with the $L$-distance from the main reflection.
As soon as the deposition begins, Laue oscillations from the  $(222)_\mathrm{Fe_3O_4}$-reflection emerge, and their period changes with growing film thickness. 
The $(224)_\mathrm{Fe_3O_4}$, however, is absent at the beginning, before it gradually appears. Nearing the end of the growth process, it also develops a Laue fringe. After the end of growth, the period of the Laue fringes remains constant, pointing to a stable film thickness and interface roughness.

In Fig.~\ref{fig_p2MgO}(b), the $L$-integrated intensities in the white boxes in Fig.~\ref{fig_p2MgO}(a) as a function of film thickness are plotted, and Fig.~\ref{fig_p2MgO}(c) shows a close-up of the temporal evolution during growth of the first $3\,$nm. Immediately after the beginning of the deposition, the intensity at the \mbox{$(2,2,2-\epsilon)_\mathrm{Fe_3O_4}$} position starts to oscillate due to the emergence of Laue fringes of the $(222)_\mathrm{Fe_3O_4}$-reflection, which are changing frequency during the growth (cf. Fig.~\ref{fig_p2MgO}(a)). With a delay of about $\Delta d=11\pm 1.3\,$\r{A}, the intensity of the $(224)_\mathrm{Fe_3O_4}$ starts to rise, very similar to the case of the $\mathrm{Fe_3O_4/SrTiO_3}$.

We used the same procedure to analyze the $\mathrm{Fe_3O_4/NiO/MgO}$ sample. The intensity of the \mbox{$(2,2,2-\epsilon)_\mathrm{Fe_3O_4}$} was analyzed at $\epsilon=0.09$. The results are shown in Figs.~\ref{fig_p2MgO}(d)-(f). In Fig.~ \ref{fig_p2MgO}(d), Laue oscillations caused by the finite thickness of the pre-deposited NiO film are visible already before the deposition of $\mathrm{Fe_3O_4}$ starts. After start of the deposition, they are superimposed by the Laue-oscillations due to the $\mathrm{Fe_3O_4}$ film, resulting in a Moir\'e pattern caused by the Laue-oscillations of the two films.
As in the case of the $\mathrm{Fe_3O_4/MgO}$ sample, the intensity at the $(2,2,2-\epsilon)_\mathrm{Fe_3O_4}$ position starts oscillating immediately after the beginning of the deposition, while the intensity of the $(224)_\mathrm{Fe_3O_4}$ reacts with a delay of about $\Delta d = 10\pm 3\,$\r{A}. %Due to the higher deposition rate and the lower time resolution available for the $\mathrm{Fe_3O_4/NiO/MgO}$, the uncertainty is higher than for the other two samples, but the result is well comparable.

\section{\label{sec:Discussion}Discussion}
%This is likely related to an island growth mode of $\mathrm{Fe_3O_4}$ on $\mathrm{SrTiO_3}$, as suggested by the SAXS measurements.
The GISAXS data of the $\mathrm{Fe_3O_4}$/SrTiO\textsubscript{3} film suggest that the film starts to grow in islands. The two distinct vertical streaks observed in the early stage of the growth (cf. Fig.~\ref{fig_p2SAXS}(a)) are an indicator for a cluster formation on the substrate surface, and their orientation parallel to the $q_\perp$ direction corresponds to the formation of islands of cylindrical or box shape \cite{Renaud09}. Their separation distance is a measure of the mean distance of the clusters, and has been observed to decrease with growing film thickness (cf. Fig. \ref{fig_p2SAXS}(c)), corresponding to a coarsening process due to, e.g., coalescence of smaller islands to bigger ones. From the film thickness at which the two streaks are not visibly separated anymore, we can conclude that a fully closed film is formed not before a coverage of $2.2\,$nm. The appearance of the second set of streaks at $2.9\,$nm, exhibiting a $55^\circ$ tilt to the out-of-plane direction (cf. Fig. \ref{fig_p2SAXS}(b)), is an indication for the formation of (111) facets \cite{Jiang14}. This means that after the first layer is closed, the film continues to grow in (001) direction, but develops (111) facets on the surface, pointing towards pyramid-shaped islands on the surface. The development of (111) facets has been reported earlier for $\mathrm{Fe_3O_4}$  films, and is related to the (111) surface being more stable than the (001) surface \cite{Takahashi14,Parkinson16_2}. 
The GISAXS data collected of the $\mathrm{Fe_3O_4}$/MgO(001) film (not shown) displays initially the same behavior: the film grows islands which coalesce to a closed layer at a coverage of about $2-3\,$nm. However, no small angle signal indicating the formation of additional nanostructures such as (111) facets is observed. This suggests that the $\mathrm{Fe_3O_4}$/MgO(001) film continues to grow in layer-by-layer mode. 

While for the two samples grown on MgO(001), the $(222)_\mathrm{Fe_3O_4}$-reflection emerges immediately after begin of the deposition, for $\mathrm{Fe_3O_4/SrTiO_3}$ the $(222)_\mathrm{Fe_3O_4}$ appears only after about $1.5\,$nm are grown already.
This implies a lower initial ordering in films below a coverage of $1.5\,$nm deposited on SrTiO\textsubscript{3}(001) than for films on MgO(001). Such a distorted interface layer of $\mathrm{Fe_3O_4}$/SrTiO\textsubscript{3}(001) has been reported before \cite{Kuschel17}.
A possible explanation is that the competition between a (001) and a (111) orientation reported for $\mathrm{Fe_3O_4}$/SrTiO\textsubscript{3}(001) \cite{Takahashi12,Takahashi14} might cause different orientations for different islands, so that no Bragg peaks develop due to missing long range order during this growth stage. However, our data do not allow a clarification of the structure in this phase.

\begin{figure}[t]
\centering
\includegraphics[scale=0.4]{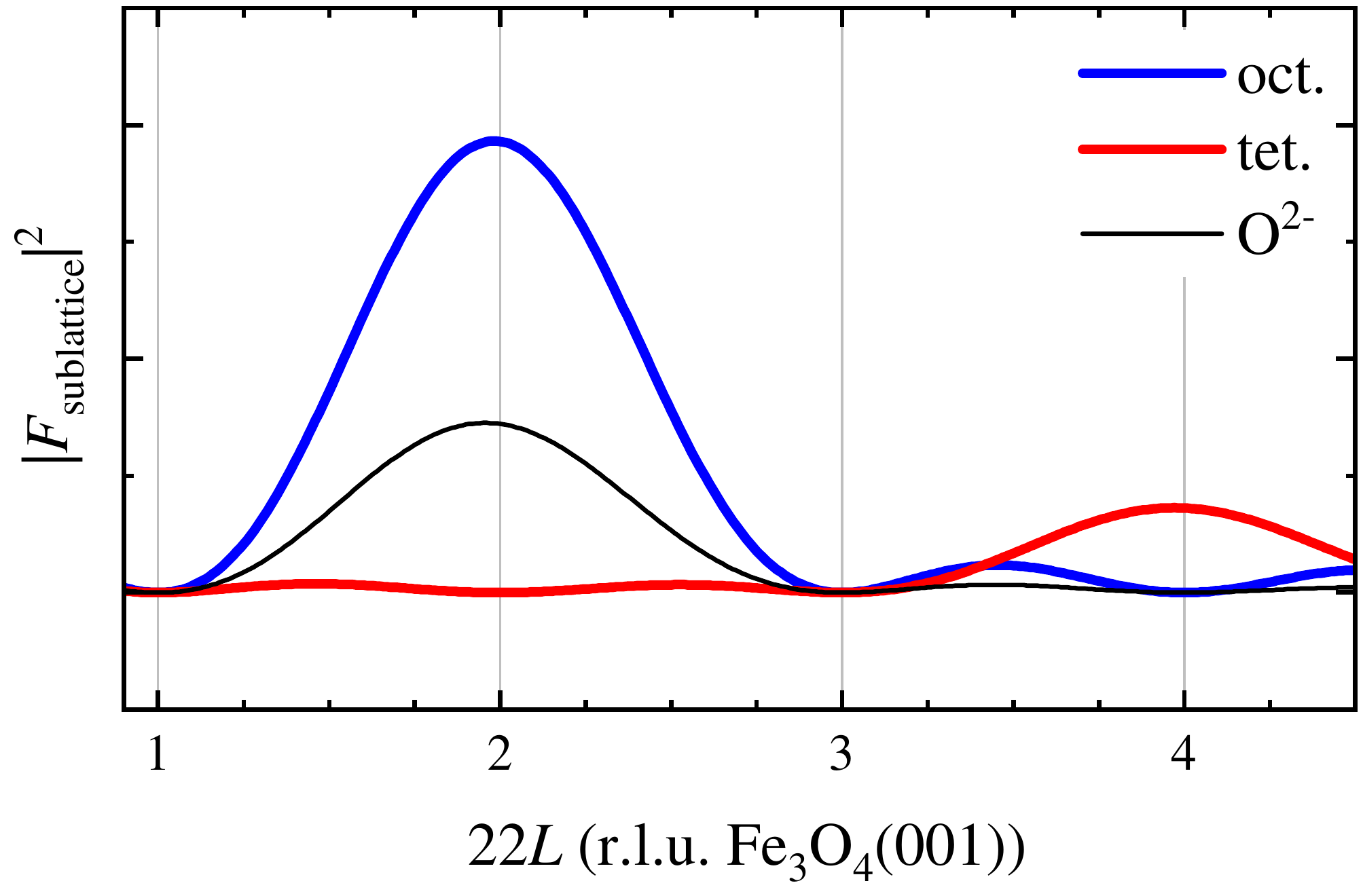}
\caption{Contributions of the octahedral, tetrahedral and oxygen sublattices to the structure factor $|$\textit{F}$_\text{sublattice}|^2$ of $\mathrm{Fe_3O_4}$.}
\label{fig_p2F}
\end{figure}

The most striking result for all three samples is the fact that after the $(222)_\mathrm{Fe_3O_4}$ reflections and their diffraction rods appear, the spinel-exclusive $(224)_\mathrm{Fe_3O_4}$ reflection follows only after the film has grown thicker by $1.1\pm 0.15\,$nm.
The inverse spinel structure of $\mathrm{Fe_3O_4}$ can be described as consisting of a cubic close-packed oxygen sublattice and two cation sublattices, one containing the octahedrally coordinated $\mathrm{Fe^{2+/3+}_{oct}}$ cations (B-sites) and one containing the tetrahedrally coordinated $\mathrm{Fe^{3+}_{tet}}$ cations (A-sites). These sublattices give rise to different Bragg-reflections, some of which are exclusive to the spinel structure while others also occur in a rock salt phase \cite{Bertram13}. 
Figure \ref{fig_p2F} shows the structure factors $|F_\mathrm{sublattice}|^2$ for the A-, B- and $\mathrm{O^{2-}}$ sublattices. The intensity of the $(222)_\mathrm{Fe_3O_4}$ has contributions from both the oxygen and the octahedral sublattice, while the intensity of the $(224)_\mathrm{Fe_3O_4}$ is purely due to the tetrahedral sublattice. Therefore, the delayed emergence of the $(224)_\mathrm{Fe_3O_4}$ with respect to the emergence of the $(222)_\mathrm{Fe_3O_4}$ indicates that in the early growth stage, the iron oxide film grows in a rock salt structure instead of the inverse spinel structure of $\mathrm{Fe_3O_4}$. 

Additionally, the position of the (222) reflection of the film agrees well with the expected position of the bulk (111)\textsubscript{Fe\textsubscript{1-$\delta$}O} reflection, and only relaxes towards the lattice constant of $\mathrm{Fe_3O_4}$  when the film grows thicker.
This suggests that iron oxide films which are grown under conditions that are well-known to grow in the $\mathrm{Fe_3O_4}$ phase \cite{Ketteler01,Celotto03,Kuschel16,Bertram13,Rodewald19} begin to grow in a Fe\textsubscript{1-$\delta$}O structure with exclusive occupation of octahedral lattice sites, and only after a certain thickness start to form the inverse spinel structure of $\mathrm{Fe_3O_4}$ with Fe cations occupying also tetrahedral lattice sites, as well. 
We determine the thickness of this rock salt layer to be about $1\,$nm, as this is the thickness difference between the emergence of the two reflections consistent for all three samples. This corresponds to about 2--3 unit cells of Fe\textsubscript{1-$\delta$}O, or about 4--6 atomic layers \cite{Bertram12}.
We also want to emphasize that this finding is robust although the growth rates differ by almost an order of magnitude between the samples (cf. Tab. \ref{tab_p2lattice}) and different substrates have been used. 
A similar effect has been reported earlier for $\mathrm{Fe_3O_4}$/MgO(001) films deposited at low temperatures or low deposition rates \cite{Bertram13}. It has been attributed to an iron-deficient rock salt structure which retains the same stoichiometry as $\mathrm{Fe_3O_4}$. 

In our tr-HAXPES data obtained for the growth of $\mathrm{Fe_3O_4}$  on SrTiO\textsubscript{3}(001), we observe a charge-transfer satellite at the high-energy side of the $\mathrm{Fe~2p_{3/2}}$ and a chemical shift of the $\mathrm{Fe~2p_{3/2}}$ and the $\mathrm{Fe~2p_{1/2}}$ to about $0.5\,$eV lower energies for iron oxide films with thicknesses between $\sim 1.5\,$nm to $\sim 4\,$nm. Both, the satellite and the chemical shift in Figs.~\ref{fig_p2HAX}(a)-(c) are indicators for an excess of $\mathrm{Fe^{2+}}$ cations. This also suggests a Fe\textsubscript{1-$\delta$}O phase, consistent with the diffraction data.
%This result may suggest a $\mathrm{Fe_{1-\delta}O}$ phase, but this is not the only possibility. For example, the grey line in Fig.~ \ref{fig_p2HAX}(b) is the sum of the $\mathrm{Fe^{2+}_{oct}}$ and the $\mathrm{Fe^{3+}_{oct}}$ spectra, representing a $\mathrm{Fe^{3+}_{tet}}$-depleted magnetite phase. This simulation exhibits a very similar satellite, and shows a chemical shift of about $0.45\,$eV, very close to the shift found in the data. 
The chemical shift and the satellite are most pronounced for the  $1.9\,$nm and the $1.7\,$nm-thick films in Figs.~\ref{fig_p2HAX}(a) and \ref{fig_p2HAX}(b), respectively. They are weakened after  the thickness is increased to $3.3\,$nm and $3.4\,$nm, and vanish at thicknesses of $3.8\,$nm and $5.1\,$nm. The information depth of our HAXPES measurements is about $\mathrm{ID(95)}=16\,$nm, and therefore accesses the entire film thickness. However, the vanishing of the spectral features does not necessarily imply a vanishing of the $\mathrm{Fe^{2+}}$-rich Fe\textsubscript{1-$\delta$}O phase itself, but can also be attributed to its signal becoming invisible due to the $\mathrm{Fe_3O_4}$ phase forming on top. 

\begin{figure}[b]
\centering
\includegraphics[scale=0.4]{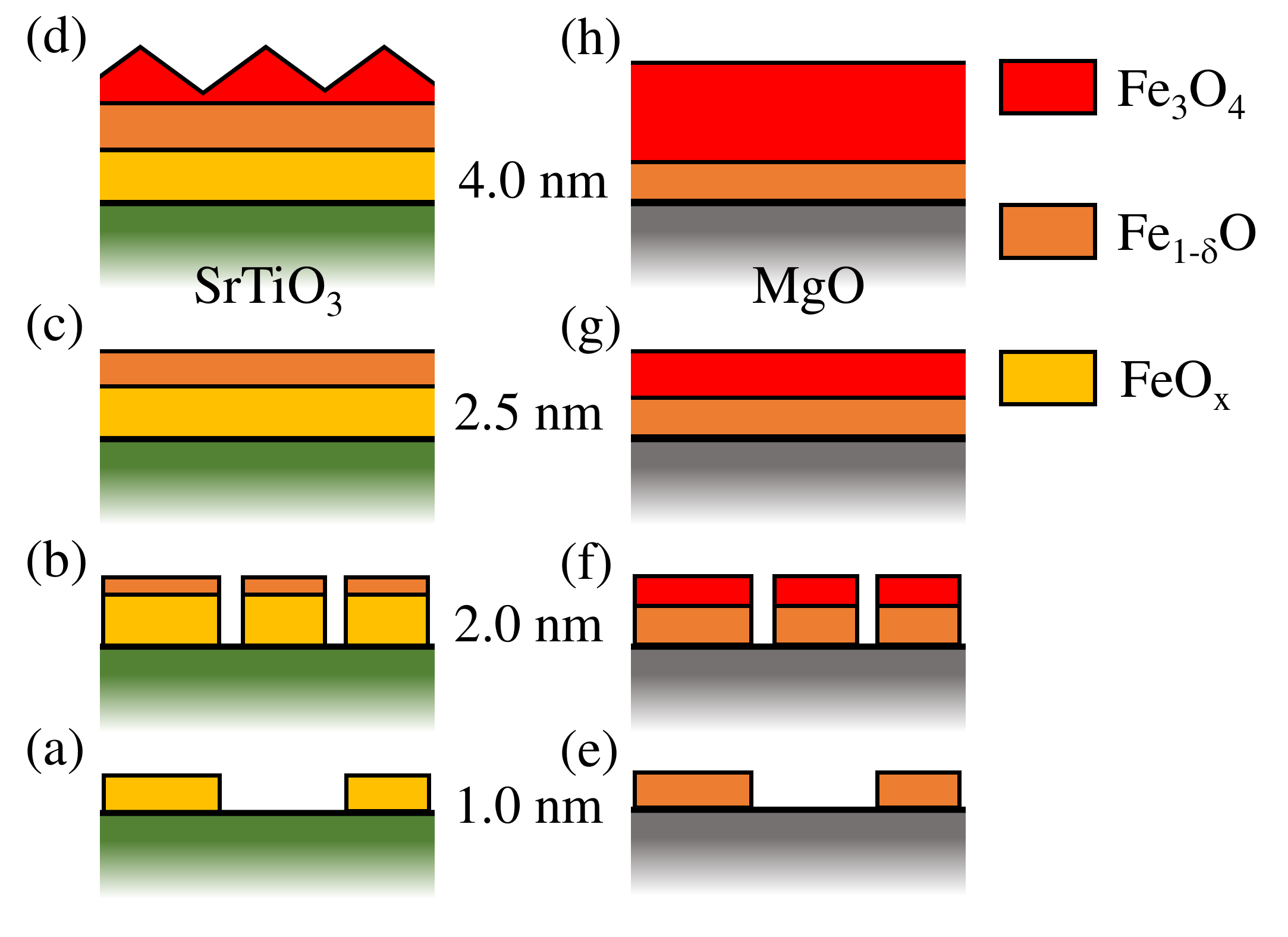}
\caption{Illustration of the growth stages at $1\,$nm, $2\,$nm, $2.5\,$nm and $4\,$nm for (a)-(d) Fe\textsubscript{3}O\textsubscript{4}/SrTiO\textsubscript{3}(001) and (e)-(h) Fe\textsubscript{3}O\textsubscript{4}/MgO(001). FeO\textsubscript{x} denotes an iron oxide phase of unclear stoichiometry.}
\label{fig_p2layers}
\end{figure} 

The tr-HAXPES data of the step-wise deposited film also reveal that the stoichiometry of the film changes when the growth is interrupted (cf. Fig.~\ref{fig_p2HAX}(b)). During deposition, the $1.7\,$nm film clearly shows a $\mathrm{Fe^{2+}}$-charge-transfer satellite, also seen for the continuously deposited sample (cf. Fig.~\ref{fig_p2HAX}(a)). However, when the iron supply is cut, the $1.7\,$nm-thick film exhibits a spectrum characteristic for $\mathrm{Fe_3O_4}$ , and as soon as the deposition is continued, the satellite returns. This suggests that this Fe\textsubscript{1-$\delta$}O phase is a transient phenomenon and exclusively occurs during the dynamic growth process, but further oxidizes to $\mathrm{Fe_3O_4}$ when the deposition is stopped.

\section{\label{sec:level6}Summary}
In summary, the initial growth stage of $\mathrm{Fe_3O_4}$/SrTiO\textsubscript{3}(001) appears to occur in 3 steps, as illustrated in Fig.~\ref{fig_p2layers}(a)-(d): First, the iron oxide film grows in disordered islands of unknown structure. Second, at a coverage of about $1.5\,$nm, as the islands become bigger, they predominantly form a rock salt structure and show an excess of $\mathrm{Fe^{2+}}$, likely being a Fe\textsubscript{1-$\delta$}O phase. Between a coverage of $2.2\,$nm and $3\,$nm, the first layer closes, and at about $2.8\,$nm the film finally starts to grow in an inverse spinel structure in [001] direction, developing (111) facets on the surface. 

The stages of $\mathrm{Fe_3O_4}$ films on MgO(001) and on NiO/MgO(001) are illustrated in Fig.~\ref{fig_p2layers}(e)-(h). They start to grow in islands, too, and the first closed layer forms at a coverage of about $2-3\,$nm. However, the immediate appearance of the $(222)_\mathrm{Fe_3O_4} $ reflection after the begin of the deposition suggests that these islands immediately exhibit a well ordered rock salt structure. After a thickness of about $1\,$nm, they grow in the inverse spinel structure. 
It seems very likely that it is due to the formation of a Fe\textsubscript{1-$\delta$}O layer before the $\mathrm{Fe_3O_4}$ phase starts to grow for all three samples.

%We used tr-HEXRD and tr-HAXPES measurements in order to monitor the growth of $\mathrm{Fe_3O_4}$ ultrathin films on $\mathrm{SrTiO_3(001)}$, $\mathrm{MgO(001)}$ and $\mathrm{NiO/MgO(001)}$ using RMBE.
%The $\mathrm{Fe_3O_4}$/SrTiO\textsubscript{3}(001) film initially forms disordered islands which merge to a closed layer at a coverage between $2-3\,$nm. At a coverage of $1.5\,$nm, the film starts to grow in a $1.2\,$nm-thick layer of $\mathrm{Fe_{1-\delta}O}$, confirmed by both its lattice constant and an excess of $\mathrm{Fe^{2+}}$ cations seen in tr-HAXPES measurements. After a film thickness of $2.9\,$nm, the film grows in the inverse spinel structure of $\mathrm{Fe_3O_4}$  and develops (111)-oriented nanofacets on the surface. 
%The films on MgO(001) and NiO/MgO(001) grow initially in an island mode as well, and form a closed layer after the deposition of about $2 - 3\,$nm. For the first $1\,$nm, the films exhibit a rock salt structure just like the $\mathrm{Fe_3O_4}$/SrTiO\textsubscript{3}(001) film, likely being a $\mathrm{Fe_{1-\delta}O}$ phase as well. After a thickness of $1\,$nm, the films grown in the inverse spinel structure of $\mathrm{Fe_3O_4}$  on both MgO(001) and NiO/MgO(001). 

The tr-HAXPES measurements of $\mathrm{Fe_3O_4}$/SrTiO\textsubscript{3}(001) additionally reveal that the Fe\textsubscript{1-$\delta$}O phase in the sub-nanometer range is only stable during the deposition process, but turns into a $\mathrm{Fe_3O_4}$  phase when the deposition is interrupted. We therefore conclude that this is a strictly dynamic property of the growth process.

\section{\label{sec:level7} Acknowledgments}
Financial support from the Bundesministerium f\"ur Bildung und Forschung (FKZ 05K16MP1) is gratefully acknowledged.  We are also grateful for the kind support from the Deutsche Forschungsgemeinschaft (DFG under  No. KU2321/6-1, and No. WO533/20-1).
We acknowledge DESY (Hamburg, Germany), a member of
the Helmholtz Association HGF, for the provision of experimental
facilities. Parts of this research were carried out at
PETRA III and we would like to thank Ren\'e Kirchhof for assistance in using P07-EH2.
\bibliography{DissBib}
\end{document}